\documentclass[aps,prl,twocolumn,showpacs]{revtex4-1}

\usepackage{graphicx}
\usepackage{dcolumn}
\usepackage{bm}
\usepackage{braket}

\begin{document}

\preprint{APS/123-QED}

\title{Multi-photon interference in the spectral domain \\by direct heralding of superposition states}

\author{Bryn A. Bell}
\email{bryn.bell@physics.ox.ac.uk}
\author{Benjamin J. Eggleton}
\affiliation{Centre for Ultrahigh bandwidth Devices for Optical Systems (CUDOS), Institute of Photonics and Optical Science (IPOS), School of Physics, University of Sydney, NSW 2006, Australia.
}

\date{\today}

\begin{abstract}
Multi-photon interference is central to photonic quantum information processing and quantum simulation, usually requiring multiple sources of non-classical light followed by a unitary transformation on their modes. Here, we observe interference in the four-photon events generated by a single silicon waveguide, where the different modes are six frequency channels. Rather than requiring a unitary transformation, the frequency correlations of the source are configured such that photons are generated in superposition states across multiple channels, and interference effects can be seen without further manipulation. This suggests joint spectral engineering is a tool for controlling complex quantum photonic states without the difficulty of implementing a large unitary interferometer, which could have practical benefits in various applications of multi-photon interference.
\end{abstract}

\maketitle
Photonic quantum information processing and quantum simulation rely on the interference of many single photons in large unitary interferometers~\cite{OBrien2007}. While universal, fault-tolerant quantum computing has a very large resource overhead, theoretical advances have inspired hope for near-term quantum photonic devices that outperform classical computers for specific tasks; so-called boson sampling with tens of photons in thousands of interfering modes is thought to be sufficient to challenge existing supercomputers~\cite{Aaronson2013, Lund2014, Neville2017}. However, experimental implementations are still far from this point - up to 5 photons in 9 modes and 3 photons in 13 modes have been demonstrated~\cite{Tillman2013, Spring2013, Broome2013, Carolan2014, Bentivega2015, Wang2017}. As the number of modes is increased, a larger number of optical elements is required to manipulate them, leading to increasing levels of loss, and higher photon number states are increasingly sensitive to loss.

There have been several proposals to take advantage of the large information capacity of a single optical fiber by interfering many temporal modes~\cite{Motes2014, He2017, Pant2016} or frequency channels~\cite{Lukens2016, Joshi2017}. These architectures can reduce the experimental complexity of the unitary interferometer, but it remains challenging to manipulate many photons with low loss and high fidelity. Boson sampling with a spatial unitary transformation and the addition of temporal or spectral correlation measurements has also been studied theoretically~\cite{Laibacher15,Tamma15, Laibacher17}.

\begin{figure}[t]
	\centering
	\includegraphics[width=\columnwidth]{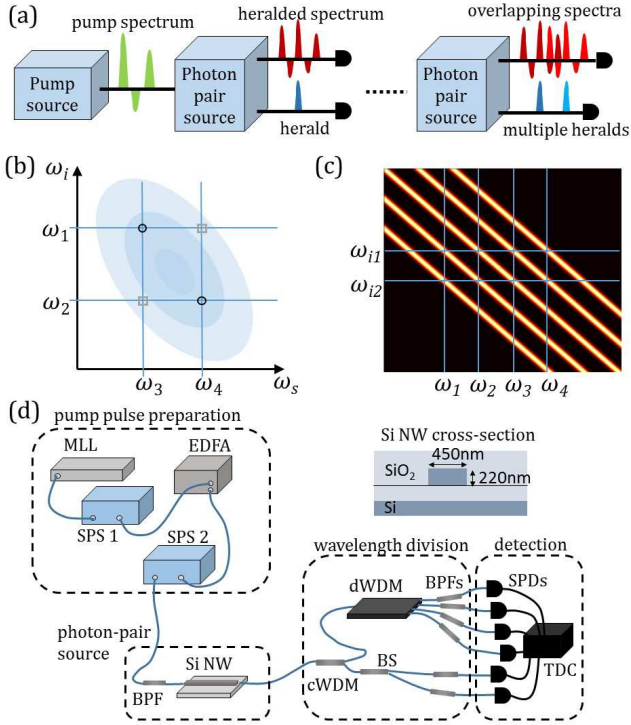}
        \vspace{-10pt}
	\caption{(a) A photon-pair source with a complex pump laser spectrum is used to herald photons in frequency superposition states. Two heralding photons close in frequency give rise to overlapping superpositions where multi-photon interference can be seen. (b) The four photon generation probability can be related to the permanent of a 2x2 matrix containing four points from the two photon JSA. (c) The experimental JSA consists of up to five diagonal lines created by FWM from different pump components. The six output frequency channels are marked with blue lines. (d) Experimental setup, consisting of pump pulse preparation using a mode-locked laser (MLL), spectral pulse shapers (SPS) and erbium-doped fiber amplifier (EDFA); photon pair generation in a silicon nanowire (SiNW), cross-section shown in inset; wavelength division (WDMs); and detection, using super-conducting nanowire single photon detectors (SPDs) and a time-to-digital converter (TDC).}\label{fig1}
 \vspace{-5pt}
\end{figure}

Here, we experimentally demonstrate multi-photon interference across frequency channels without applying a unitary transformation after the quantum light source. Instead, the frequency correlations, or joint spectral amplitude (JSA), of the source are configured such that photons are directly generated in superpostion states across several channels [Fig.\ref{fig1}(a)]. Then, when the photon number in each frequency channel is sampled, multi-photon interference effects can be seen without the need for further manipulation, and without the complexity and associated loss of a multi-mode unitary transformation. The quantum light is generated by spontaneous four-wave mixing (FWM) in a silicon nanowire (SiNW)~\cite{Sharping06}, and a non-trivial JSA is created by shaping the complex envelope of the pump pulses~\cite{Grice97, Brecht2015, Ansari2017}. After the source, six frequency channels are monitored with single photon detectors, and interference is observed in the different combinations of four photon events between these channels.


Figure.~\ref{fig1}(b) depicts a general JSA where $\omega_s$ and $\omega_i$ denote the frequencies of signal and idler photons. The probability of four photons being created at the four marked frequencies $\omega_{1-4}$ is:
\begin{equation}
P\propto|\psi_{1,3}\psi_{2,4}+\psi_{1,4}\psi_{2,3}|^2,
\end{equation}
where $\psi_{j,k}$ is the amplitude for creation of a pair of photons at frequencies labeled $j$ and $k$. There are two separate paths to creating this four photon state which can combine coherently and interfere, corresponding to different permutations of signal and idler pairings. This is equal to the permanent of a 2x2 matrix containing the four amplitudes. For higher photon numbers, the probabilities of $N$-pair generation would depend on permanents of $N$x$N$ matrices~\cite{supp}, analogous to the output probabilities in boson sampling, which depend on matrix permanents of input-output transition amplitudes~\cite{Aaronson2013}.

The SiNW photon-pair source has a large phase-matched bandwidth of around 20THz at telecommunication wavelengths, so over the range of frequencies used the JSA is determined only by energy-matching - the sum of the energies of the two pump photons involved in the FWM process must be equal to the sum of the signal and idler photons. For a monochromatic pump, the JSA is a diagonal line such that signal and idler must be equally spaced about the pump frequency~\cite{Jizan2015}. Here, the pump pulses consist of multiple frequency components, creating a JSA containing multiple diagonal lines, as shown in Fig.~\ref{fig1}(c). Detecting an idler photon at a particular frequency heralds a signal photon in a superposition of multiple possible frequencies (or vice versa). The separations of the output frequency channels (four for the signal and two for the idler) are matched to the separations of the pump frequency components, as in Fig.~\ref{fig1}(c). This results in 8 possible two photon outputs containing one signal and one idler, and 6 possible four photon outputs containing two signals and two idlers (we do not use photon number resolving detection, so events containing two photons in the same channel are not recorded).

The experimental setup is shown in Fig.~\ref{fig1}(d). To prepare multiple pump frequencies which are phase-stable with respect to each other, a mode-locked laser (MLL) with 25nm bandwidth around 1552nm is filtered into separate channels by a spectral pulse-shaper (SPS, Finisar Waveshaper). The pump channels are centered around 1550.5nm with 200GHz separation, and individual bandwidths of 50GHz. The pump pulses are amplified by an erbium-doped fiber amplifier (EDFA), then noise from the spontaneous emission of the EDFA is removed using a second SPS and a band-pass filter (BPF).

\begin{figure}[b]
	\centering
	\includegraphics[width=\columnwidth]{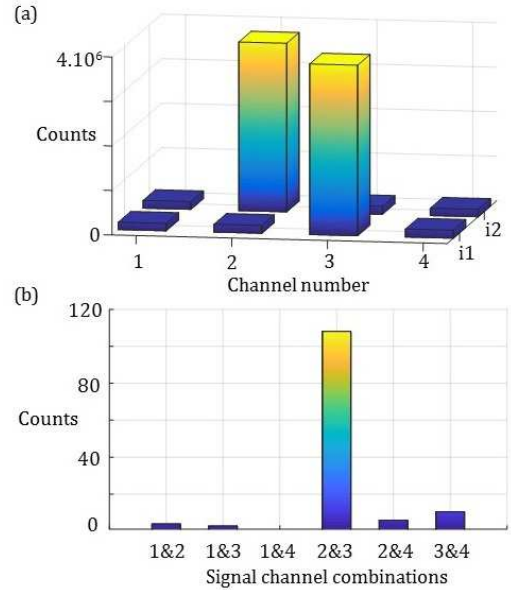}
        \vspace{-10pt}
	\caption{(a) Single pump component, two photon counts, showing that idler channel i1 (i2) heralds signal channels 2 (3), according to energy-matching. (b) Four-photon counts involving both idler channels and two out of four signal channels. Signal photons in channels 2$\&$3 are heralded.}\label{fig2}
 \vspace{-5pt}
\end{figure}

The SiNW has cross-sectional dimensions of 450nm$\times$220nm and a length of 3mm. The insertion loss is 6dB, of which around 5dB is due to mode mis-match when coupling to and from fiber. After the SiNW, a course wavelength division multiplexer (cWDM) is used to separate the generated idler photons at $>1565$nm from the signal and pump. The two idler channels are then separated by a beam-splitter (BS) followed by BPFs tuned to 1568.7nm and 1570.4nm, with bandwidths of 0.6nm (75GHz). The four signal channels are separated by a dense wavelength division multiplexer (dWDM) - the wavelengths were 1529.6nm, 1531.1nm, 1532.7nm, and 1534.3nm, with channel bandwidths of 0.4nm (50GHz). Tunable BPFs on each channel further suppressed the bright pump light. Each channel was sent to a superconducting nanowire single photon detector (SPD), and the times of the detections were recorded using a time-to-digital converter (TDC). Due to variation in the filter transmissions and detector efficiencies, with overall heralding efficiencies in the range $2-4\%$, all count rates were normalized to the least efficient signal and idler channels in post-processing.

Initially, a single pump component was used at 1550.5nm. This results in a matrix of two photon amplitudes~\cite{supp}:
\begin{equation}
\psi\propto\left(\begin{array}{cccc}
{0} & {1} & {0} & {0} \\
{0} & {0} & {1} & {0} 
\end{array}\right).
\end{equation}
The corresponding two photon detections in a one hour integration time are shown in Fig.~\ref{fig2}(a). Detecting an idler photon in channel i1 (i2) heralds a signal in channel 2 (3). Fig.~\ref{fig2}(b) shows the 6 combinations of four photon counts, each of which involves detection of an idler photons in i1 and i2, which heralds signal photons in both channels 2\&3. A relatively high multi-photon emission probability results in some other channel combinations, particularly channels 3\&4. In this case, the photons are heralded in particular channels, not superposition states, so no interference is expected.

\begin{figure}[t]
	\centering
	\includegraphics[width=\columnwidth]{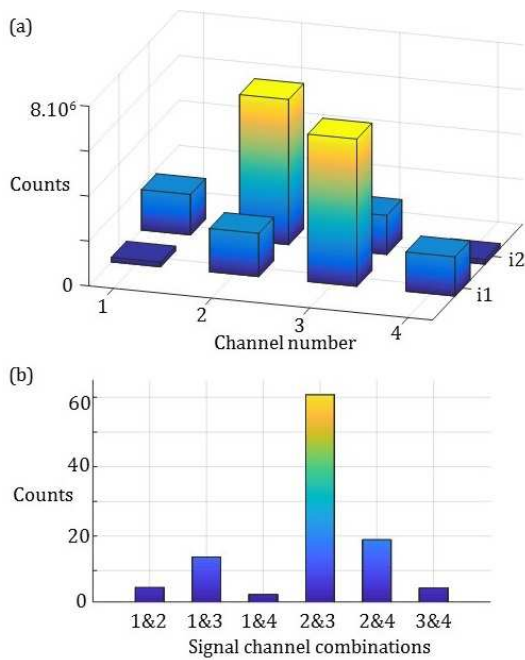}
        \vspace{-10pt}
	\caption{(a) Two pump components, two photon counts; i1 and i2 now herald overlapping superposition states across the signal channels. (b) Four photon counts; multi-photon interference is expected to increase the 2$\&$3 combination slightly.}\label{fig3}
 \vspace{-5pt}
\end{figure}

Next, two equal intensity pump components were used, centered around 1550.5nm. The pump power was decreased and the integration time increased to 8 hours, to reduce the multi-photon noise. The two photon amplitudes are now:
\begin{equation}
\psi\propto\left(\begin{array}{cccc}
{1} & {2} & {1} & {0} \\
{0} & {1} & {2} & {1} 
\end{array}\right).
\end{equation}
There are three entries in each row, corresponding to different FWM processes. The entries with amplitude 1 correspond to FWM from the individual pump components, and the entries with amplitude 2 correspond to the brighter non-degenerate FWM process involving both pumps~\cite{supp}. Fig.~\ref{fig3}(a) shows the corresponding two photon events - note that the count rates depend on $|\psi_{i,j}|^2$, so the central peaks are expected to be four times larger than the side peaks. The four photon counts are shown in Fig.~\ref{fig3}(b). All combinations of signal channels are expected to contain some events because choosing any two columns from $\psi$ forms a 2x2 matrix with non-zero permanent, but the combination 2$\&$3 still dominates, with 1$\&$3 and 2$\&$4 the next most significant.

The 2$\&$3 combination depends on two non-zero permutations of signal and idler pairings which should add constructively. We can compare the measured four photon counts to those expected based on statistical combinations of the two photon counts. In the absence of quantum interference (ie. if there is distinguishing information to say which signal photon is paired with which idler photon), the event probability $P(\vec{j}, \vec{k})=\text{perm}(|\psi_{\vec{j}, \vec{k}}|^2)$ can be calculated directly from the two photon count rates~\cite{Tichy2015}. The measured counts are greater than those predicted in the absence of quantum interference (61$\pm$8 compared to 45), suggesting constructive interference takes place, but the effect is not very noticeable because the interfering paths are highly unbalanced in amplitude, and could conceivably be attributed to statistical uncertainty. We note that this is not affected by the phases of the two pumps, so it is not possible to tune between constructive and destructive interference~\cite{supp}.

\begin{figure}[b]
	\centering
	\includegraphics[width=\columnwidth]{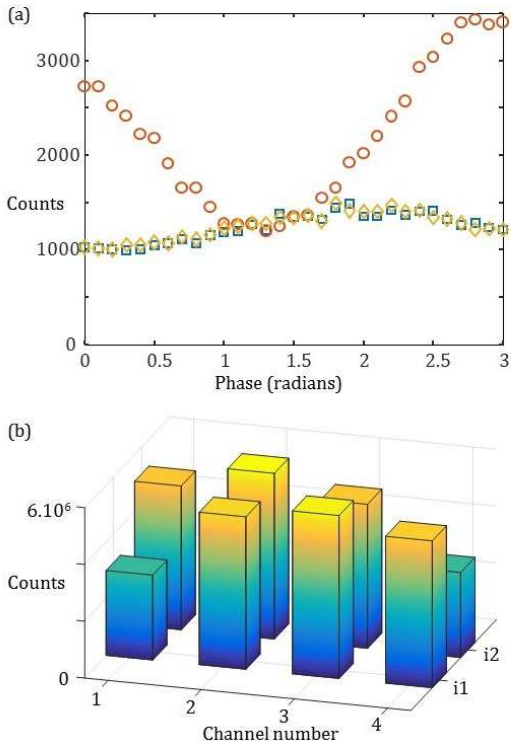}
        \vspace{-10pt}
	\caption{(a) Three pump components; variation in two photon count rates as a function of the phase of the center pump. Blue squares: i1$\&$1; Red circles: i1$\&$2; Yellow diamonds: i1$\&$3. (b) All two photon combinations with the phase fixed at 1.3 rad. i1 and i2 now herald highly overlapping superpositions across the signal channels, and some amplitudes carry a $\pi/2$ phase-shift not apparent here.}\label{fig4}
 \vspace{-5pt}
\end{figure}

Tuning the complex amplitudes of three pump components provides greater flexibility to reconfigure the JSA. The two photon amplitude matrix can be written
\begin{equation}
\psi\propto\left(\begin{array}{llll}
{2A_1A_2~} & {A_2^2+2A_1A_3~} & {2A_2A_3} & {A_3^2} \\
{A_1^2} & {2A_1A_2} & {A_2^2+2A_1A_3~} & {2A_2A_3}
\end{array}\right),
\end{equation}
where $A_{1-3}$ are the complex amplitude of the three pumps~\cite{supp}. Note that now two processes contribute to the entries $\psi_{i1,2}$ and $\psi_{i2,3}$ - FWM from $A_2$, and non-degenerate FWM involving both $A_1$ and $A_3$ - and so these amplitudes are sensitive to the pump phases. We aim to set $A_1=A_3=\sqrt{2}iA_2$, which gives
\begin{equation}
\psi\propto\left(\begin{array}{cccc}
{2\sqrt{2}i} & {-3} & {2\sqrt{2}i} & {-2} \\
{-2} & {2\sqrt{2}i} & {-3} & {2\sqrt{2}i}
\end{array}\right),
\end{equation}
providing relatively balanced amplitudes and a mix of phases which can lead to both constructive and destructive multi-photon interference.

We use the second SPS in the setup to attenuate $A_2$ relative to $A_1$ and $A_3$, and to apply a phase-shift. Fig.~\ref{fig4}(a) shows the variation in three of the two photon count rates as this phase is changed. As expected, the count rate corresponding to $\psi_{i1,2}$ is highly dependent on the phase, but unexpectedly some slower variation can be seen in the count rates corresponding to $\psi_{i1,1}$ and $\psi_{i1,3}$. This is potentially explained by the transfer of energy between the pump components due to phase-sensitive parametric amplification in the SiNW, with energy either moving from the central to the outer pump components or vice versa depending on the phases~\cite{Zhang2014}. Despite this, setting the phase of the central pump to 1.3 radians appears to give the desired configuration; the fact that $|\psi_{i1,3}|^2$ is minimized suggests the phase is set correctly, and the count rates are reasonably balanced. Fig.~\ref{fig4}(b) shows all of the two photon counts observed in a 12 hour integration time.

\begin{figure}[t]
	\centering
	\includegraphics[width=\columnwidth]{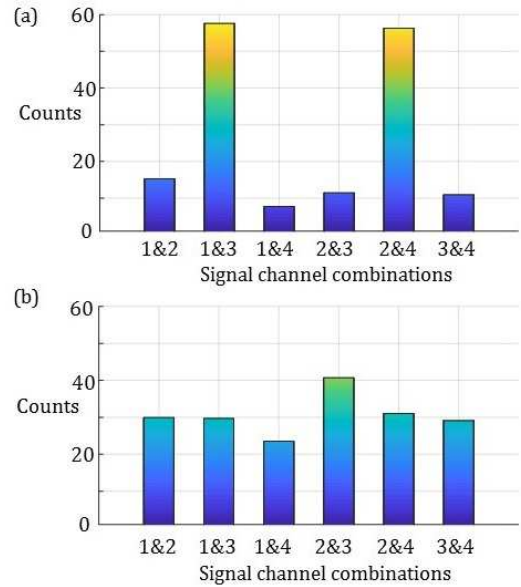}
        \vspace{-10pt}
	\caption{(a) Four photon counts, displaying constructive interference for signal channels 2$\&$4 and 1$\&$3, but destructive interference in the other combinations. (b) These features are absent in the expected counts without multi-photon interference, inferred from the two photon counts.}\label{fig5}
 \vspace{-5pt}
\end{figure}

Figure~\ref{fig5}(a) shows the measured four photon counts, and Fig.~\ref{fig5}(b) shows the four photon counts expected from statistical combinations of the two photon events. A pattern can be seen in the measured counts which is not predicted by classical statistics. This can be understood in terms of the permanents of 2x2 matrices derived from $\psi$. Choosing the 1st and 3rd columns, the contributions from the two possible permutations add constructively, leading to an increased count rate for the 1$\&$3 combination of signal channels. The same is true for choosing the 2nd and 4th columns, and the 2$\&$4 combination of signal channels. For all other combinations of columns corresponding to signal channels, there is a destructive interference effect which reduces the corresponding count rate.

While this measurement gives a clear indication of multi-photon interference effects occurring, the interference contrast here is imperfect - on average the count rates exhibiting constructive interference are 5 times larger than those exhibiting destructive interference. We note that perfect contrast is not expected, due to the imbalances in the interfering amplitudes. Higher-order photon emission ($>4$) contributes noise which will also tend to reduce the contrast. Finally, in order to see high-contrast interference, it is necessary for the channel filters to be narrow compared to the bandwidth of the individual pump components, so that post-filtering all spectral correlation between photons is removed. Unfortunately narrowing the filters sacrifices heralding efficiency and makes multi-photon experiments impractical~\cite{MeyerScott2017}. This could be improved by placing the source inside a cavity, since then the photons are created in narrow resonance modes without the need for narrow filtering~\cite{Clemmen2009, Chen2011, Azzini2012, Reimer2016,Kues2017}.

Scaling this scheme up entails adding more frequency channels, detecting higher photon numbers, and ideally adding more control over the JSA function. Adding more frequency channels is relatively straight-forward, especially compared to adding more modes to a spatial interferometer, because the phase-matched bandwidth of the FWM is large and many-channel telecommunication filters are readily available. It is also possible to monitor many frequency channels with a single SPD using a single photon spectrometer in the style of~\cite{Avenhaus2009}.

Detecting higher photon numbers requires reducing the loss experienced by the generated photons. Here, we were limited to four photon events with low count rates, largely due to lossy filtering, and an immediate improvement could be seen by replacing the tunable band-pass filters with low-loss fixed filters~\cite{Zhang2017}. The SiNW itself has relatively high propagation and coupling losses and could be replaced by another broadband photon-pair source~\cite{Dayan2004, Lim2008}. As discussed above, making use of a cavity-based source provides an improved trade-off between heralding efficiency and interference contrast.

Here, the form of the JSA is limited to functions of the form $f(\omega_s+\omega_i)$, whereas for quantum information applications one would ideally have universal control of the JSA function. While this is likely challenging, strategies for more general manipulation of the JSA include dispersion-engineering to tailor the phase-matching of the source~\cite{Mosley2008} and using silicon photonic resonantors or band-gap structures to modify the JSA~\cite{Kumar2014, Helt2017}. For parametric downconversion with $\chi^{(2)}$ crystals, domain engineering could be used to tailor the output~\cite{URen2006, Titchener17}.

In summary, we have applied pulse-shaping to the complex envelope of the pump laser prior to generating photon-pairs by broadband FWM in a SiNW. This results in a complex JSA which exhibits quantum interference in the four photon emission, because there is no distinguishing information to say which signal photon was created in a pair with which idler photon - the interfering paths correspond to different pairings of the generated photons. Compared to experiments which use multiple sources of quantum light followed by a unitary interferometer, this avoids much of the experimental complexity, and so could have practical benefits for many applications of quantum interference and multi-photon states.

\begin{acknowledgments}
This work was supported by the Australian Research Council (ARC) Centre of Excellence CUDOS (CE110001018) and Laureate Fellowship (FL120100029).
\end{acknowledgments}

\vspace{2cm}
\onecolumngrid
\section{\fontsize{18}{25}\selectfont Supplementary Information}
\section{Multi-photon coincidence amplitudes}

For convenience, we will assume the JSA is a matrix of amplitudes corresponding to photon-pair creation at particular signal and idler frequencies. Since the JSA is actually a continuous function, the bandwidth of the frequency channels should be small compared to the scale over which it varies, otherwise detecting a photon in a particular channel can project it onto a mixed spectral state. In the experiment, the photons are split into relatively broad frequency channels and the resulting mixing degrades the interference visibility. On the other hand, if the JSA were to already consist of a discrete set of modes, for instance the modes of a cavity-based photon pair source, the filtering would only be required to separate these distinct modes. 

Creation operators for the signal channels are labeled by $a^{\dagger}_j$ and for the idler channels $b^{\dagger}_j$. A JSA can generally be decomposed into pairs of Schmidt modes~\cite{McKinstrie}. The Schmidt modes provide an orthonormal basis such that each pair of modes is an independent two-mode squeezed state, and the overall wavefunction is the tensor product of these squeezed states. We use $c^{\dagger}_j$ for the creation operators of the signal Schmidt modes, and $d^{\dagger}_j$ for the idler Schmidt modes. The two-mode squeezed-vacuum state between the $j$th pair of Schmidt modes can be written as:
\begin{equation}
\sqrt{1-\lambda_j^2}\sum_{n=0}^\infty \lambda_j^n\ket{n,n}_{j,j}=\sqrt{1-\lambda_j^2}~\text{exp}\left(\lambda_j c^{\dagger}_jd^{\dagger}_j\right)\ket{0,0}_{j,j}.
\end{equation}
Where $\ket{n,m}_{j,k}$ indicates a number state with $n$ photons in the $j$th signal Schmidt mode and $m$ photons in the $k$th idler Schmidt mode. $\lambda_j$ is a parameter describing the strength of the squeezing. Hence the total state is
\begin{equation}
\prod_j \sqrt{1-\lambda_j^2}~\text{exp}\left(\lambda_j c^{\dagger}_jd^{\dagger}_j\right) \ket{\text{vac}}=\mathcal{C}~\text{exp}\left(\sum_j\lambda_j c^{\dagger}_jd^{\dagger}_j\right)\ket{\text{vac}}=\mathcal{C}~\text{exp}\left(\sum_{j,k}\Lambda_{j,k} a^{\dagger}_jb^{\dagger}_k\right)\ket{\text{vac}}
\end{equation}
where $\ket{\text{vac}}$ is the multimode vacuum state, $\Lambda$ is a matrix describing the strength of squeezing between different signal and idler modes in the frequency basis, and $\mathcal{C}=\prod_j \sqrt{1-\lambda_j^2}$. We have used the fact that the $c^{\dagger}_j$ ($d^{\dagger}_j$) operators are just linear combinations of the $a^{\dagger}_j$ ($b^{\dagger}_j$) operators. It can be seen that the Schmidt basis puts $\Lambda$ into diagonal form and is helpful in determining the constant $\mathcal{C}$. The exponential can be expressed as a series of terms corresponding to creation of different total photon numbers:
\begin{equation}
\text{exp}\left(\sum_{j,k}\Lambda_{j,k} a^{\dagger}_jb^{\dagger}_k\right)=1+\left(\sum_{j,k}\Lambda_{j,k} a^{\dagger}_jb^{\dagger}_k\right)+\frac{1}{2}\left(\sum_{j,k}\Lambda_{j,k} a^{\dagger}_jb^{\dagger}_k\right)^2+...\frac{1}{N!}\left(\sum_{j,k}\Lambda_{j,k} a^{\dagger}_jb^{\dagger}_k\right)^N
\end{equation}
eg. the two photon amplitude associated with one signal photon in mode $j$ and an idler in mode $k$ is $\mathcal{C}.\Lambda_{j,k}$. The amplitude associated with $N$ signal photons in modes $\vec{j}=j_1,...,j_N$ and $N$ idler photons in modes $\vec{k}=k_1,...,k_N$ depends only on the $N$th term of the expansion:
\begin{equation}
\phi(\vec{j}, \vec{k})=\frac{\mathcal{C}}{N!}\bra{\text{vac}}\left(\prod_{p=1}^N a_{j_p}\right)\left(\prod_{q=1}^N b_{k_q}\right)\left(\sum_{r,s}\Lambda_{r,s} a^{\dagger}_rb^{\dagger}_s\right)^N\ket{\text{vac}}.
\end{equation}
First we observe that the $N$ different $b_{k_q}$ must match up with the $N$ different $b_s^\dagger$ to give a non-zero amplitude. There are $N!$ possible orderings which give the same result, so the pre-factor of $1/N!$ is cancelled out:
\begin{equation}
\phi(\vec{j}, \vec{k})=\mathcal{C}\bra{\text{vac}}\left(\prod_{p=1}^N a_{j_p}\right)\left(\prod_{q=1}^N\sum_r\Lambda_{r,k_q}~a^{\dagger}_r\right)\ket{\text{vac}}.
\end{equation}
Now the $N$ different $a_{j_p}$ must match the $a_r^\dagger$, but this time the ordering matters, because different orderings give rise to different pairings of indices in the $N$ instances of the $\Lambda$ matrix. Hence there is a summation over all permutations of $j_1...j_N$:
\begin{equation}
\phi(\vec{j}, \vec{k})=\mathcal{C}\sum_{\sigma\in S_N}\prod_{q=1}^N \Lambda_{j_{\sigma(q)}, k_q}=\mathcal{C}~\text{perm}(\Lambda_{\vec{j},\vec{k}}).
\end{equation}
Where $S_N$ is the set of all permutations of the numbers 1 to $N$. This is equal to the permanent of $\Lambda_{\vec{j},\vec{k}}$, a sub-matrix of $\Lambda$ containing the rows and columns corresponding to the generated signal and idler frequencies, multiplied by the constant $\mathcal{C}$. The summation is essentially over different pairings of the $N$ signal photons with the $N$ idler photons.

In the main text, we refer to the permanent of a submatrix $\psi_{\vec{j}, \vec{k}}$ which contains the two photon amplitudes. As seen above, the two photon amplitudes are equal to the elements of $\Lambda$ multiplied by $\mathcal{C}$. Hence the permanents derived from these two photon amplitudes have an extra factor of $\mathcal{C}^N$, and
\begin{equation}
\phi(\vec{j}, \vec{k})=\mathcal{C}^{1-N}~\text{perm}\left(\psi_{\vec{j}, \vec{k}}\right)
\end{equation}
\begin{equation}
P(\vec{j}, \vec{k})=\mathcal{C}^{2-2N}~|\text{perm}\left(\psi_{\vec{j}, \vec{k}}\right)|^2.
\end{equation}
So the probabilities are equal to the absolute square of the permanent, up to a constant factor, which is equal to 1 in the limit of small squeezing.

\section{JSA matrices for experimental pump configurations}

A continuous JSA $\psi(\omega_s, \omega_i)$ determined only by the energy matching condition for FWM can be written:
\begin{equation}
\psi(\omega_s, \omega_i)\propto\int d\omega~e(\omega)~e(\omega_s+\omega_i-\omega)=f(\omega_s+\omega_i)
\end{equation}
where $e(\omega)$ is the spectral amplitude of the pump pulse, $\omega-\omega_s-\omega_i$ is the frequency of the second pump photon, fixed by energy matching, and this results in a function $f(\omega_s+\omega_i)$ which only depends on the sum of the signal and idler frequencies. For a discretized JSA, $\psi_{j,k}$, and discrete pump components with amplitudes $e_j$, where the subscripts label equally spaced frequency channels, we can write
\begin{equation}
\psi_{j,k}\propto\sum_l e_l~e_{j+k-l}=f_{j+k}.
\end{equation}

For a single pump component in channel $p$, $\psi_{j,k}$ is only non-zero when $j+k=2p$. So fixing the idler channel, there is only one non-zero component for the signal, and displacing the idler by one channel must also displace the signal by one channel. Hence we can write
\begin{equation}
\psi\propto\left(\begin{array}{cccc}
{0} & {e_p^2} & {0} & {0} \\
{0} & {0} & {e_p^2} & {0} 
\end{array}\right),
\end{equation}
where the two rows correspond to the two idler channels used in the experiment, and the four columns correspond to the four signal channels.

For two pump components with amplitudes $e_p$,$e_q$, the pump function is
\begin{equation}
f_j=\sum_k e_k e_{j-k}~~\Rightarrow~~f_{2p}=e_p^2,~~f_{p+q}=2e_pe_q,~~f_{2q}=e_q^2,
\end{equation}
so if $p$ and $q$ are adjacent channels we have
\begin{equation}
\psi\propto\left(\begin{array}{cccc}
{e_p^2} & {2e_pe_q} & {e_q^2} & {0} \\
{0} & {e_p^2} & {2e_pe_q} & {e_q^2} 
\end{array}\right).
\end{equation}
There are two paths to generating the outcome involving photons in both idler channels and signal photons in channels 2 and 3:
\begin{equation}
\phi_{i1,i2,2,3}\propto\text{perm}\left(\begin{array}{cc}
{2e_pe_q} & {e_q^2} \\
{e_p^2} & {2e_pe_q} 
\end{array}\right)=4e_p^2e_q^2+e_p^2e_q^2=5e_p^2e_q^2,
\end{equation}
where it can be seen that the two possibilities always add constructively, regardless of the complex phases of $e_p$ and $e_q$, and that they are unbalanced by a ratio 4:1, which makes this effect harder to observe definitively in an experiment.

With three pump components $e_p$,$e_q$,$e_r$, the pump function has 6 non-zero components:
\begin{equation}
f_{2p}=e_p^2,~~f_{2q}=e_q^2,~~f_{2r}=e_r^2,~~f_{p+q}=2e_pe_q,~~f_{p+r}=2e_pe_r,~~f_{q+r}=2e_qe_r.
\end{equation}
If $p$,$q$,$r$ are three adjacent channels, $f_{2q}$ coincides with $f_{p+r}$, and becomes $e_q^2+2e_pe_r$. This means an idler in a particular channel heralds a signal in a superposition of five frequencies, one of which is cut off since we only use four signal channels. This configuration provides sufficient degrees of freedom to observe both constructive and destructive interference:
\begin{equation}
\psi\propto\left(\begin{array}{llll}
{2e_pe_q~} & {e_q^2+2e_pe_r~} & {2e_qe_r} & {e_r^2} \\
{e_p^2} & {2e_pe_q} & {e_q^2+2e_pe_r~} & {2e_qe_r}
\end{array}\right).
\end{equation}

\end{document}